\documentclass[conference]{IEEEtran}
\IEEEoverridecommandlockouts
\usepackage{cite}
\usepackage{amsmath,amssymb,amsfonts}
\usepackage{algorithmic}
\usepackage{graphicx}
\usepackage{textcomp}
\usepackage{subcaption}
\usepackage{multirow}
\usepackage{tabularx}

\def\BibTeX{{\rm B\kern-.05em{\sc i\kern-.025em b}\kern-.08em
    T\kern-.1667em\lower.7ex\hbox{E}\kern-.125emX}}
\begin{document}

\title{Cooling Down FaaS: \\
      Towards Getting Rid of Warm Starts
}

\author{\IEEEauthorblockN{D\'aniel G\'ehberger}
\IEEEauthorblockA{\textit{Ericsson Research} \\
Montreal, Canada \\
daniel.gehberger@ericsson.com}
\and
\IEEEauthorblockN{D\'avid Kov\'acs}
\IEEEauthorblockA{\textit{Ericsson Research} \\
Budapest, Hungary \\
kovacs.david@ericsson.com}
}

\maketitle

\begin{abstract}
Serverless execution and most notably the Function as a Service model got quite some attention during the recent years. As of today, all commercial and open source implementations follow the common practice of keeping the execution environments running to achieve low function execution latency. In this paper we compare the startup latency of different available virtualization technologies, then we implement and benchmark an FaaS prototype system using IncludeOS unikernels for function execution. We show that our system can start and execute functions with essentially the same latency as AWS Lambda with its continuously running executor units. Due to the low overhead, this approach opens the possibility for simplified FaaS platforms without the resource waste and extensive monitoring requirements of existing solutions.
\end{abstract}

\begin{IEEEkeywords}
cloud, serverless, FaaS, unikernel
\end{IEEEkeywords}

\section{Introduction}

Cloud technologies have been continuously evolving during the recent years. One important aspect of this evolution is the increased decouplement of the applications from the underlying infrastructure. Serverless computing is the most recent step on this journey, making applications independent even from the virtual infrastructure. In practice, this means that the infrastructure scaling and maintenance tasks are all offloaded to the provider, and the developers using the platform can focus on the application logic~\cite{berkeley}.

The Function as a Service (FaaS) model is a realization of serverless computing that was first widely introduced by Amazon Web Services (AWS) through its Lambda service. Simply stated, using the FaaS development model, an application developer can write functions in any of the languages supported by the given platform, functions can be attached to event sources, or triggers and the platform automatically executes them on demand. Existing, especially commercial, platforms offer a wide-range of triggers including HTTP requests, database or other storage data updates and various message queues. The functions in FaaS are in general required to be stateless, specifically the state should be provided as input or externalized to a database.

After the introduction and initial success of AWS Lambda, other cloud players also introduced similar services. Then, following the commercial trends, open source implementations also started to be available. While they are becoming more and more mature, the stability and supported features of these options are still behind the commercial alternatives. A notable exception is OpenWhisk that is used in IBM's commercial offering. Other major open source implementations include Kubeless, OpenFaaS, Nuclio, Fission and the Fn project.

As we will show in the following sections, the virtualization technologies used in today's FaaS solutions require at least a couple of hundred of milliseconds to start. This extra startup latency is usually referred to as \textit{cold start} and directly impacts the service quality. To tackle this problem, FaaS systems use pre-warmed execution units, meaning that they keep environments up and running for a while in order to make subsequent \textit{warm starts} fast. While this approach makes cold starts rare, they still occur from time to time without the possibility to predict from the user's point of view.

A well performing FaaS system with only cold starts would be an important advancement in this technology domain. On the one hand, keeping idle environments running wastes resources, on the other hand, significant part of the complexity in existing platforms comes from the handling of warm environments, including per-function load monitoring, scaling and routing requests to proper warm environments. If a platform would be able to start functions fast enough for each individual incoming request, the system could be greatly simplified, as function scaling can be simply driven by the actual load.

In this paper, we investigate if it is possible to create such an FaaS framework. First, we dive into the recent advancements of the container technology and show its current distance from a cold-only FaaS platform, then, we turn our focus to unikernels due to their lightweight characteristics. The idea of using unikernels in FaaS has been proposed in a few research papers during the last year \cite{nabla-paper, sand, berkeley}, but to the best of our knowledge we are the first who present a working prototype.

The rest of the paper is structured as follows: in Section~\ref{sec:background} we discuss the important aspects of the technology domain, then, in Section~\ref{sec:performance} we show our startup measurement results. In Section~\ref{sec:fn} we show how unikernels can be applied in the Fn FaaS platform. Finally, we discuss the related work in Section~\ref{sec:related}, then conclude the paper.

\section{Technology Background}\label{sec:background}

In theory, fundamentally different technologies can be used to build an FaaS system, ranging from starting or forking a process through different container technologies to unikernels or complete virtual machines. In the following we discuss the possible alternatives on this scale, then in the next section we will compare their startup performance.

\subsection{About using processes} \label{sec:processes}

The most lightweight option for creating a new executor entity is by using \textit{fork()} or \textit{clone()} in Linux. According to our experiences, forking can take between 55--500~$\mu$s depending on how much memory needs to be replicated, even with Linux’s copy-on-write memory sharing. In order to have a process that can be forked for each incoming request, the process - with the function loaded - needs to be started first. As a result, the performance of starting a processes draws a baseline for cold starts.

The main limitation of processes is that while it is possible to restrict their access capabilities regarding the filesystem, networking, etc., once all needed features are enabled the system basically ends up using something like a Docker container. We will measure the time required to start different types of processes in the next section and we argue that this approach is a viable option for single-tenant, performance oriented FaaS platform. However, in this paper we seek for more isolated options from security perspective, that are confidentially applicable for multi-tenant environments.

 \subsection{Containers} \label{sec:containers}

Since its introduction in 2013 Docker became the de-facto solution for light-weight virtualization by combining several components in the kernel for separation of container instances. Due to the light-weight and high-granularity nature of FaaS systems, Docker is in general a good fit and it is used in all existing open source FaaS implementations as an execution engine.

Docker is built using several layers and starting a container requires gRPC based communication throughout its stack that includes the CLI, Docker Engine, containerd, a shim layer to decouple containers, and finally an Open Container Initiative (OCI) compatible runtime~\cite{oci}.

OCI was established in 2015 to create open industry standards for container environments. It currently maintains two specifications: the Runtime Specification and the Image Specification. The former outlines how to run a filesystem bundle or image that is unpacked on disk. In reality, the definitions are mostly based on Docker components and the runtime reference implementation, called as runc, also comes from Docker. However, since the introduction of OCI, several other runtimes have appeared, including for example Kata Containers~\cite{kata} and gVisor~\cite{gvisor}.

The true power of OCI is that the runtimes can be fundamentally different, while runc uses Linux namespaces and cgroups to create containers, gVisor is a user-space kernel that is built using syscall interception as the core component and finally Kata Containers uses Qemu with KVM to launch lightweight general purpose virtual machines and as a result blurs the distinction between containers and virtual machines.

\subsection{Virtual machines}

In theory it would be possible to use traditional virtual machines as FaaS execution units, but we ruled out this option as such a machine takes 10s of seconds to start.

AWS recently open sourced its light-weight hypervisor Firecracker that is claimed to be the backend for its Lambda and Fargate services~\cite{firecracker}. Firecracker makes use of KVM to launch micro-virtual machines, combines the security benefits of virtual machines and the resource efficiency of containers. In the next section we will show that while Firecracker is faster than Qemu, it cannot beat runc and gVisor.

The most light-weight available virtual machine options are unikernels. A unikernel is a single-purpose virtual machine, a single image including only the relevant drivers, operating system components and the application itself. They are usually single threaded and come with a single address space. Unikernels have been around for many years and due to their single-purpose nature the related solutions are mostly designed for packet processing and high-performance computing. The key advantage of unikernels is the combination of high-performance, hardware level separation and low footprint. These are made possible due to the highly-specialized nature that comes with internal simplicity.

The unikernel concept is not new and it has quite a few realizations. Most of the implementations use one of the generic, well-known hypervisors, like Xen or Qemu-KVM. In this paper we focus on IncludeOS~\cite{includeos-paper} which is a single task system written in C++. IncludeOS builds on virtio drivers, comes with its own networking stack and standard library implementation. We use IncludesOS because besides supporting Qemu-KVM it can be also compiled for solo5.

Solo5 is a sandboxed execution environment primarily intended for unikernels, thus providing extremely fast startup time~\cite{ukvm}. IncludeOS uses the \textit{hardware virtualized tender} (hvt) of solo5, that is previously known as ukvm, and builds on top of KVM. Solo5 was recently extended with a \textit{sandboxed process tender} (spt) that uses seccomp for separating processes~\cite{nabla-paper}.

In an FaaS system that uses cold-only functions the image size is an important factor, as images should be transferred and cached on a lot, in an extreme setting on all, the machines in the cluster. Looking at the different options, the solo5 example applications take only around 200~kB disk space. A simple echo server built using IncludeOS is around 2.5~MB, while a base Alpine Linux container is around 6~MB. Finally, the base Firecracker kernel is around 20~MB and the rootfs we use is around 50~MB.

\section{Comparing startup performance} \label{sec:performance}

In this section we compare the startup performance of the different virtualization technologies to get a clear picture about the different options.

\subsection{General FaaS architecture}

FaaS systems are built using highly similar main components, namely a gateway, an HTTP or event router that is also called dispatcher, a cluster manager and the function executor units. A request to run a function is received by the gateway, that passes it to the dispatcher, the dispatcher looks for available (warm) units to execute the request and may request a new, cold, unit from the cluster manager. In production ready FaaS frameworks the dispatcher also performs authentication and authorization before executing requests.

\subsection{Our measurement system}

In order to better understand the startup time overhead of the different runtime options we created a benchmarking tool using the C++ based CppCMS web framework~\cite{cppcms}. In our setup the CppCMS acts as an FaaS gateway and our measurement application running over CppCMS acts as event router and dispatcher by exposing different URLs, like \textit{/docker\_runc} or \textit{/includeOS}. On receiving an HTTP request the application executes a simple echo application using the given technology. For example if \textit{/docker\_runc} is queried, the framework starts an Alpine Docker container with \textit{/bin/date} as the command. The scaling inside our physical measurement machine is automatically handled by the CppCMS framework, which is configured to have multiple processes for accepting connections and 20 worker threads.

We use the hey HTTP load generator tool to generate requests and measure the latency~\cite{hey}. Hey is an easy to use tool and can be used to send  the given number of HTTP requests with the defined parallelism. We use a different machine to run hey, and the 2 servers are connected through a dedicated 40~Gbps Mellanox network. We are using Ubuntu 18.04 with 4.18 kernel version and our servers are equipped with Intel Xeon E5-2670 CPUs, 64~GB of memory and Samsung PM1633a SSD drives.

For the container related measurements we used Docker 18.09.3, runc 1.0.0-rc6, Kata Containers 1.4.3 and a gVisor commit from 12th of March 2019. We used Firecracker 0.15 and version 0.4 of the solo5 hypervisor for the spt measurements. Finally, we used IncludeOS 0.14 that builds on a solo5 commit from July 2018.

We used different parallelism configurations for the measurement, for example \textit{10 parallel calls} in the figures implies that 10 requests are in-flight at any given time. As our measurement machine has 24 cores, we will show the behaviour under overload conditions with setting the highest load to 40 parallel requests. We also validated that the machine sending the requests is not a bottleneck in this range. For each measurement we used 10000 requests and we use boxplots with whiskers reaching the 1st and 99th percentiles.

\subsection{Baseline Docker results}

According to our measurements, starting a single Alpine Linux Docker container via the CLI can take around 650~ms with the default runc OCI runtime in interactive mode, and 450~ms as a daemon. Starting runc directly with the most basic configuration and the exported Alpine image takes around 150~ms.

The difference is actually the accumulated overhead of various factors. In order to run an OCI runtime, a \textit{rootfs} containing the filesystem, and a \textit{configuration file} are required. Adding the namespace configurations used by Docker to the basic runc configuration file adds roughly 100~ms to the time required to start the environment. The largest overhead comes from networking configuration, followed by the mount and inter process communication namespaces.

Other than the overhead of gRPC communication throughout the Docker software stack, most of the remaining difference comes from the Docker storage drivers that are needed to create a rootfs for runc. Docker uses the \textit{overlay2} storage driver by default that is a union filesystem making it possible to place a container specific writeable layer on top of multiple read only container image layers and logically showing it as a single filesystem. We compared the different available storage drivers and found that the default option performs the best from the perspective of startup latency.

\subsection{Comparing OCI runtimes}

\begin{figure}
\includegraphics[width=0.48\textwidth]{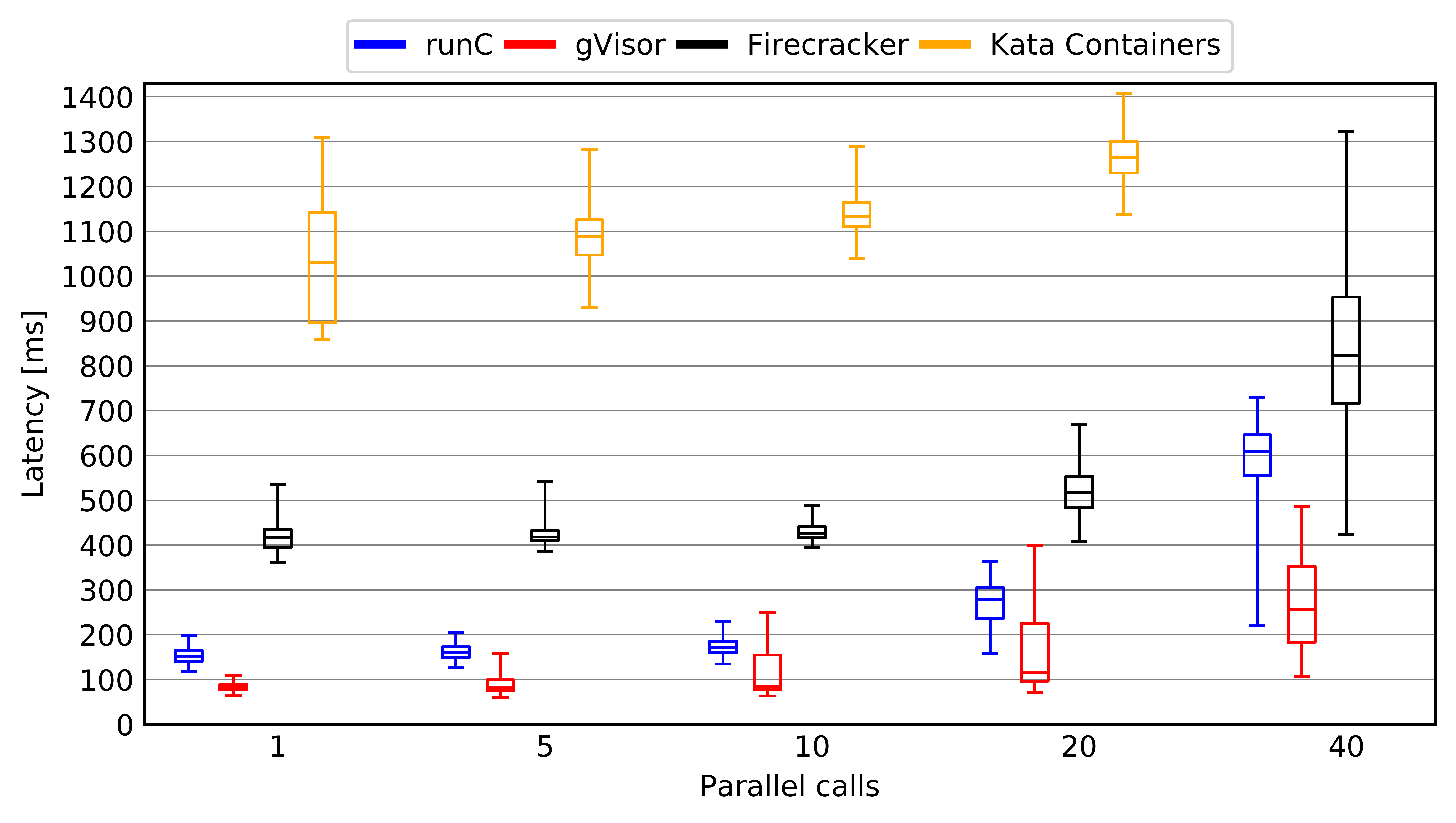}
\caption{Startup times with OCI runtimes and Firecracker. For better visibility Kata Containers is omitted under the overload condition, with median value at 2.2 seconds and 99th percentile at 3.3 seconds.}
\label{fig:OCI}
\end{figure}

\begin{figure}
\includegraphics[width=0.48\textwidth]{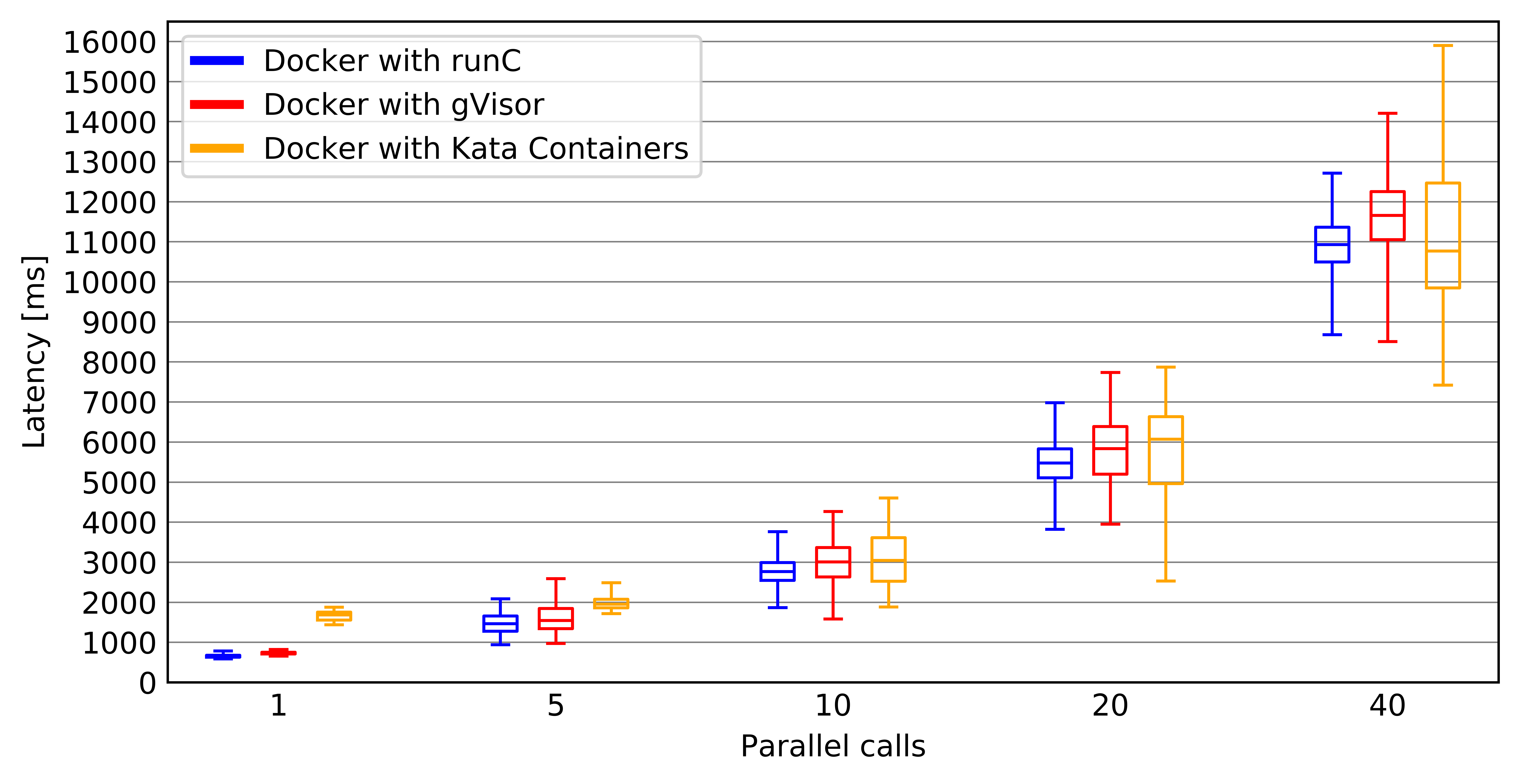}
\caption{Startup times with Docker}
\label{fig:docker_alt}
\end{figure}

The most important advantage of OCI is that it makes it possible to use radically different runtimes under Docker. First in Figure~\ref{fig:OCI} we show our measurement results under different levels of parallelism with 3 OCI runtimes, and we also include measurements with Firecracker. We use the same Alpine rootfs for all OCI runtimes, and an Alpine image for Firecracker.

As it can be seen, gVisor provides better results compared to runc, while Kata Containers is clearly slower than the other options due to the overhead of starting up Qemu-KVM each time. We also found the startup performance of Firecracker to be a quite comparable option to OCI runtimes. While all options scale fairly well up until 20 parallel start requests, when we go over the number of cores available in the server, the latency gets impacted, especially for Kata Containers.

Figure~\ref{fig:docker_alt} suggests that the overhead of the Docker layers over the OCI level, hide most of the performance differences. Moreover, starting up a container takes over 10 seconds under the highest measured load, most probably due to limitations in accessing kernel resources and creating the union filesystems.

In summary, we argue that while new OCI runtimes, such as gVisor, provide better startup characteristics, until Docker and all the required kernel configurations add hundreds of milliseconds of overhead it is not practical to build an FaaS system without warm starts using Docker.

\subsection{Unikernels and processes}

\begin{figure}[b]
\centerline{\includegraphics[width=0.48\textwidth]{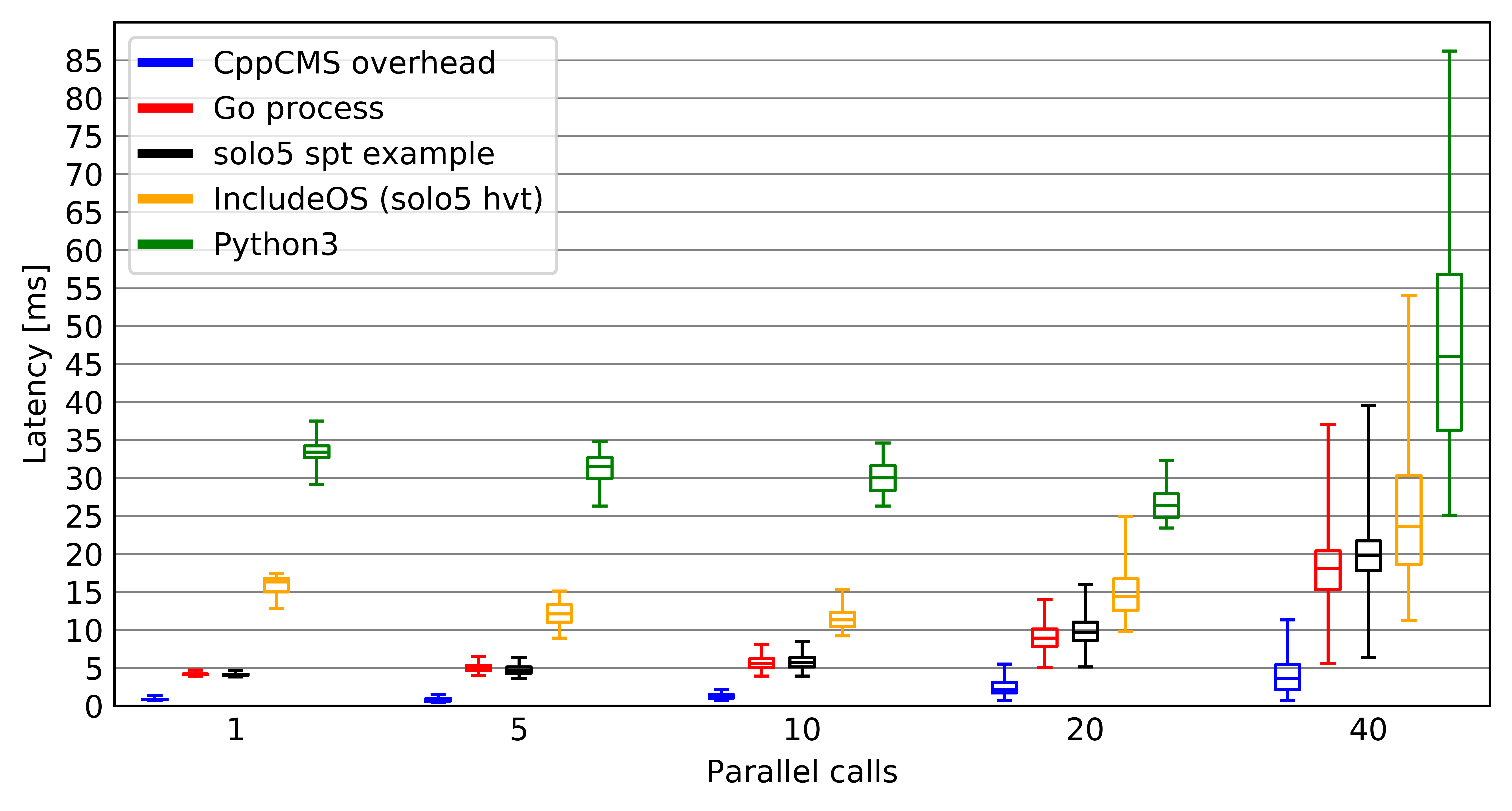}}
\caption{Measured startup times with processes and unikernels}
\label{fig:unikernels}
\end{figure}

While containers are too slow in their current form to make an FaaS platform possible without cold starts, Figure~\ref{fig:unikernels} shows that processes and unikernels provide a good opportunity. As it can be expected starting a process (e.g. a compiled Go application) brings the best latency characteristics. An interpreted language, like Python, takes significantly more time to start even without libraries. Loading a module like \textit{scipy} adds an additional 80~ms to the startup according to our experiences.

The figure shows that IncludeOS unikernel instances using solo5's hvt can start in around 8--15~ms under moderate load. We also included the basic test application of the new spt tender of solo5 and as it can be seen it gives almost the same performance as processes. The example application lacks the libraries, dynamic memory management and other features that come with IncludeOS. Once the unikernel will support this tender, the related startup times are expected to be better than with hvt.

Finally, we also measured the overhead of the CppCMS framework by adding a \textit{/noop} URL. The overhead is only around 0.7~ms for low-load scenarios, but grows considerable over 20 parallel requests. While this type of overhead is independent from the virtualization technologies, it exists in all FaaS implementations as requests need to go through the gateway and dispatcher components.

\section{FaaS with unikernels} \label{sec:fn}

Based on the results we determined that unikernels, and in our case IncludeOS specifically, can be potentially used as an FaaS runtime environment. As we will show in this section, having an FaaS with only cold IncludeOS execution is a viable alternative of keeping warm execution units.

One of the most important advantage of our proposed solution is that it does not use idle executors and thus eliminates resource waste. The idle timeout for warm executor units is obviously a configurable parameter either by the operator or the developer, however, this configuration presents a trade-off between wasting resources and experiencing frequent cold starts.

Wang et al.~\cite{curtains} analyzed FaaS platforms in public clouds and found that in the AWS cloud the Firecracker instances serving the same function are co-located to the same machine roughly while they fit into the physical memory. They measured that AWS keeps idle function executors up and running for nearly half an hour, effectively wasting significant amount of memory and CPU resources. They also reported that co-location influences startup times when sudden scale-out is required, similarly as we have shown in Figure~\ref{fig:OCI}. Compared to AWS Lambda, in Fn by default the idle timeout is configurable per function and the system keeps the executor containers in paused state, still reserving resources.

In relation the the presented alternatives, our unikernel based Fn extension essentially does not waste resources as the unikernel exits immediately after executing the user's code. Furthermore, while in a traditional FaaS platform the user can never know if a given request will experience an extended delay due to a cold start, with our solution the execution latency is predictable.

\subsection{The architecture of Fn with unikernels}

There are multiple open source FaaS solutions available, the ones with tight Kubernetes integration cannot be easily modified to run unikernels instead of Docker containers. Out of the few remaining options we choose the Go based Fn Project~\cite{fn}.

The Fn server can be separated to three main components, the \textit{gateway}, \textit{agent} and the \textit{driver}. The gateway is responsible for receiving the requests from the clients. The agent manages the life-cycle of function runtimes on the given host through the driver that handles runtime specific commands. Fn has by default only the Docker driver as fully functional implementation. We added a new driver to provide the IncludeOS support. As our unikernels exit after execution, the lifecycle management functionality of the agent becomes unnecessary with our approach.

Early versions of Fn were able to run warm and cold-only functions as well, but the former has been removed from recent releases. Currently an extra wrapper, called as FDK (Function Development Kit), is used to turn any user function into an executable container. The Docker driver communicates with the FDK running in a container via HTTP over a Unix socket, and the FDK calls the user function internally.

In our current IncludeOS driver implementation, we do not use an FDK, but use the standard in and out for communication with the unikernel, as it was done in Fn before the introduction of the FDK.

Functions can be added to the system using the \textit{deploy} command of the Fn CLI. With Docker containers, the user function and a \textit{yaml} configuration file are required, and the CLI tool will automatically create a Docker container by wrapping the function using the proper language specific FDK. Following this approach, we added an extra option to indicate that an IncludeOS build is required for the given function. The \textit{boot} IncludeOS build script is then used to create the solo5 specific image that is placed to a specific directory on the host.

When a function is called, the new driver starts the deployed IncludeOS image using the solo5 hypervisor, gives the received user input as parameter and waits for output on the stdout. After the execution of the function, the unikernel simply exits, and, in parallel, the user gets back the result.

\subsection{Measurement results}

To present our solution in a realistic configuration, we deployed our modified Fn platform into the Stockholm region of the AWS cloud. The deployment contained both the original Docker and our IncludeOS based Fn versions. We used an \textit{m5.metal} instance for the deployment as IncludeOS needs access to KVM that is only available with \textit{metal} instances. We used Postgress as the backend database for Fn as we got significant performance improvements compared to the default sqlite option. We also deployed a Go Lambda function into the same region and made it accessible through the AWS API Gateway.

We performed measurements from Ericsson's lab in Stockholm and summarized the results in Table~\ref{tab:inaws}. As it can be seen, the unikernel based approach gives an order of magnitude better cold start time compared to both the unmodified Fn and AWS Lambda.

\begin{table}[ht]
\centering
\begin{tabular}{| l || c | c | c |}
\hline
\multicolumn{1}{|c||}{Environment} & Cold start & Warm start & Connection setup \\
\hline\hline
Fn IncludeOS & 33.4 & - & \multirow{2}{*}{6.9} \\
\cline{1-3}
Fn Docker & 288.3 & 13.6 & \\
\hline
AWS Lambda &  449.7 & 78.0 & 50.1 \\
\hline
\end{tabular}
\caption{Median function execution latency measured from Stockholm with Fn and Lambda deployed in the AWS Stockholm region. The numbers are in ms.}
\label{tab:inaws}
\end{table}

An important difference between the Fn and the Lambda deployments is that the API Gateway uses TLS, that adds considerable overhead to the connection setup time due to the required 3 round-trips and the computational costs. In the table we included the connection setup both for cold and warm starts, also meaning that re-using the same TCP/TLS connection (if possible) is a powerful optimization option. Our solution with cold starts gives roughly the same latency as warm functions in AWS Lambda, considering the connection creation overhead.

For comparison, using an EC2 instance in the same AWS region as a measurement point gives only slightly lower connection setup overhead. As expected, the overhead grows with distance, getting up to around 200~ms if the Lambda function is called from our lab in Budapest.

To evaluate the pure, worst-case difference between the Docker and IncludeOS based Fn solutions, we ran a set of measurements in our local lab environment. First, we observed a notable difference in the deployment time, the C++ compilation in case of IncludeOS takes about 3.5 seconds, while Docker requires 9-10 seconds to create the image.

\begin{figure}[t]
\centerline{\includegraphics[width=0.48\textwidth]{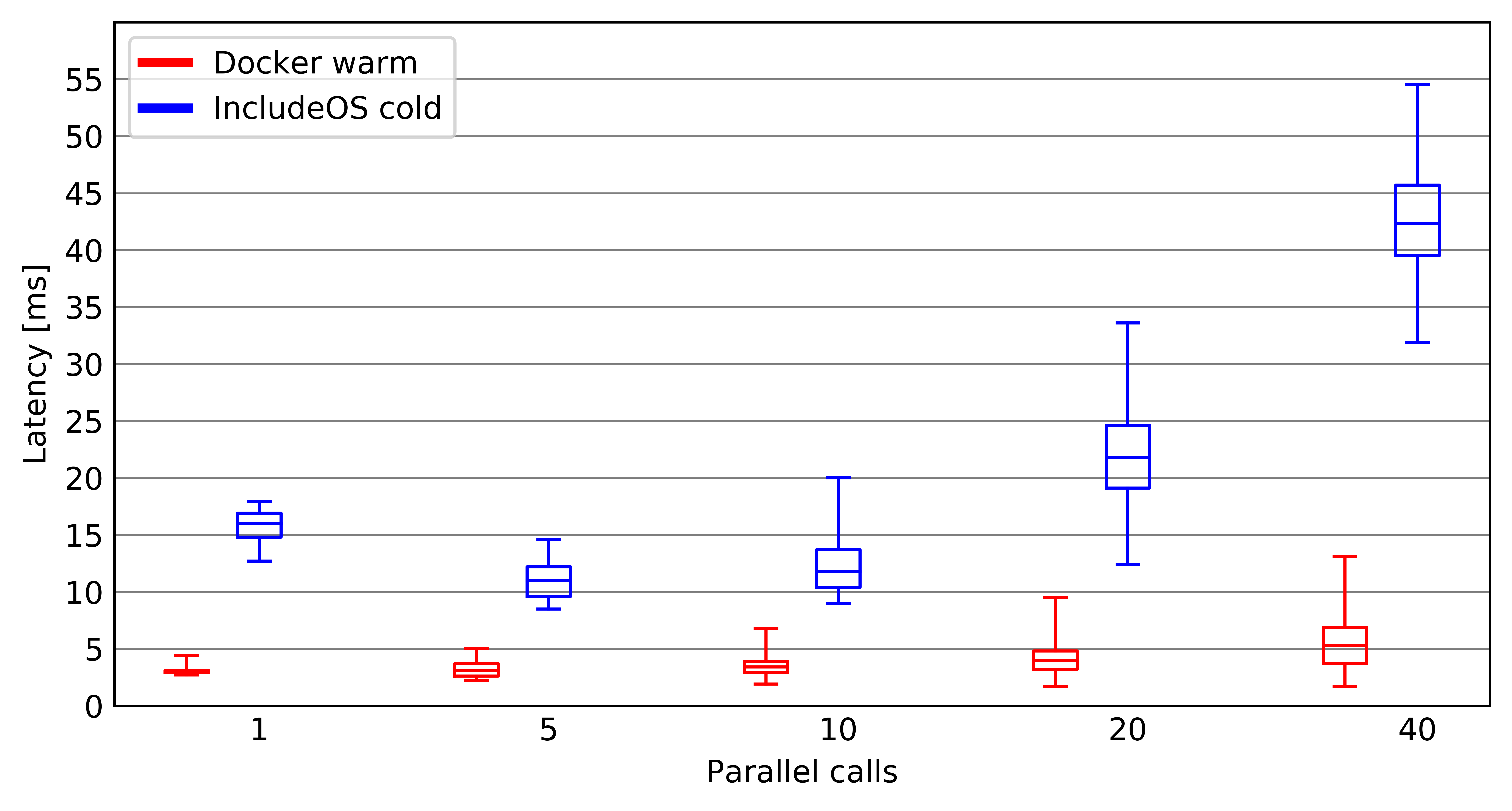}}
\caption{Fn measurement results in our local lab environment}
\label{fig:fn}
\end{figure}

Figure~\ref{fig:fn} shows that the startup and execution of our test function with IncludeOS takes around 10-20~ms. In comparison, the latency with a warm Go function takes 3-5~ms, with the price of wasting the resources reserved by the continuously running Docker containers when they are idle even for a few milliseconds.

Finally, we must note that by using more complex functions the overhead of Fn with IncludeOS gets less and less significant compared to the execution time. This minimal overhead enables creating FaaS environments with cold starts only, without the need to continuously monitor, scale and run idle function executor units, everything can be scheduled on demand.

\subsection{Limitations}

In this section we showed that our unikernel based FaaS approach provides similar latency characteristics compared to traditional warm execution approaches. Our solution has a few limitations in its current form that require further work. Probably the most interesting question is the proper handling and distribution of function images. While Docker has a complete ecosystem for container image management, there is no standard solution available for unikernels at the moment. Furthermore, we get the best performance if the image is available on the machine that gets the request.

As we discussed before, starting an interpreted language like Python with complex modules needed for the function adds around 80~ms overhead to the execution. Due to this limitation the presented unikernel approach in its current form is better suited for compiled languages like C++ or Go.

\section{Related Work} \label{sec:related}

Serverless technologies and especially FaaS got a lot of attention during the recent years. Various papers evaluated the performance, usability and security aspects of both commercial \cite{curtains, cold-factors, performance-factors} and open source \cite{open-serverless-eval, open_source_review} FaaS options.

Jonas et al. highlighted in their technical report that performance, especially predictable performance, is a key limitations in today's solutions \cite{berkeley}. They also suggested that demand is increasing for fine grained security contexts and mentioned unikernels as a possible solution to minimize the attack surface.

Besides deployment and security aspects, Wang et al. analyzed the cold start performance of FaaS implementations in the AWS, Azure and Google clouds~\cite{curtains}. They found that, depending on the configuration, the cold starts take a few hundred milliseconds in AWS and Google clouds and around 3.5~seconds in Azure with high fluctuation in the latter.

Manner et al. investigated multiple factors influencing cold starts in the AWS and Azure clouds~\cite{cold-factors}. They found that cold starting Java based functions takes 2-3 times more than using JavaScript for the same purpose, with the latter taking around 600~ms in the best scenarios.

Akkus et al. suggested in their paper that unikernels are a viable option for serverless, with concerns about limitation on flexibility~\cite{sand}. In their solution they addressed cold starts by using long running Docker containers to separate users and internally forked a worker process for each request.

The development of solo5's spt was done through the work on Nabla containers~\cite{nabla-paper}. Nabla is an OCI compliant runtime that can only run special container images with a binary for the solo5 spt inside. While our measurements show that solo5 spt provides extraordinary startup times, adding Docker on top of it basically hides its advantages compared to our proposed solution with IncludeOS.

\section{Conclusion}

In this paper we showed that using unikernels as the runtime technology is a feasible option for FaaS systems. We demonstrated that while the container technology advances quickly, the startup still takes at least a few hundred milliseconds making warm execution units necessary. We showed that unikernels provide a good alternative with startup times under 15~ms, enabling FaaS platforms without all the resource waste and complexity needed for keeping the environments warm.


\begin{thebibliography}{00}
\bibitem{sand} Istemi Ekin Akkus et al. “SAND: Towards High-Performance Serverless Computing”. In: 2018 USENIX Annual Technical Conference (USENIX ATC 18). Boston, MA, 2018, pp. 923–935..
\bibitem{firecracker} Jeff Barr. Firecracker - Lightweight Virtualization for Serverless Computing. Tech. rep. Amazon Web Services Inc., Nov. 2018. URL: https://aws.amazon.com/blogs/aws/firecracker-lightweight-virtualization-for-serverless-computing/.
\bibitem{includeos-paper} A. Bratterud et al. “IncludeOS: A Minimal, Resource Efficient Unikernel for Cloud Services”. In: 2015 IEEE 7th International Conference on Cloud Computing Technology and Science. Nov. 2015, pp. 250–257.
\bibitem{cppcms} CppCMS. http://cppcms.com/. Accessed: 2019-03-30.
\bibitem{fn} Fn Project. https://fnproject.io/. Accessed: 2019-03-30.
\bibitem{gvisor} gVisor Project. https://gvisor.dev/. Accessed: 2019-04-29.
\bibitem{hey} Hey. https://github.com/rakyll/hey. Accessed: 2019-03-30.

\bibitem{berkeley} Eric Jonas et al. Cloud Programming Simplified: A Berkeley View on Serverless Computing. Tech. rep. UCB/EECS-2019-3. EECS Department, University of California, Berkeley, Feb. 2019.
\bibitem{kata} Kata Containers Project. https : / / katacontainers . io/. Accessed: 2019-04-29.
\bibitem{open_source_review} Kyriakos Kritikos and Pawel Skrzypek. “A Review of Serverless Frameworks”. In: 2018 IEEE/ACM International Conference on Utility and Cloud Computing Companion, UCC Companion 2018, Zurich, Switzerland, December 17-20, 2018. 2018, pp. 161–168.
\bibitem{performance-factors} W. Lloyd et al. “Serverless Computing: An Investigation of Factors Influencing Microservice Performance”. In: 2018 IEEE International Conference on Cloud Engineering (IC2E). Apr. 2018, pp. 159–169.

\bibitem{cold-factors} J. Manner et al. “Cold Start Influencing Factors in Function as a Service”. In: 2018 IEEE/ACM International Conference on Utility and Cloud Computing Companion (UCC Companion). Dec. 2018, pp. 181–188.
\bibitem{open-serverless-eval} Sunil Kumar Mohanty, Gopika Premsankar, and Mario Di Francesco. “An Evaluation of Open Source Serverless Computing Frameworks”. In: 2018 IEEE International Conference on Cloud Computing Technology and Science (CloudCom) (2018), pp. 115–120.
\bibitem{oci} OCI. https://www.opencontainers.org/. Accessed: 2019-04-29.
\bibitem{curtains} Liang Wang et al. “Peeking Behind the Curtains of Serverless Platforms”. In: 2018 USENIX Annual Technical Conference (USENIX ATC 18). Boston, MA, 2018, pp. 133–146.
\bibitem{ukvm} Dan Williams and Ricardo Koller. “Unikernel Monitors: Extending Minimalism Outside of the Box”. In: 8th USENIX Workshop on Hot Topics in Cloud Computing (HotCloud 16). Denver, CO, 2016.
\bibitem{nabla-paper} Dan Williams et al. “Unikernels As Processes”. In: Proceedings of the ACM Symposium on Cloud Computing. SoCC ’18. Carlsbad, CA, USA, 2018, pp. 199–211.


\end{thebibliography}

\end{document}